\documentclass[preprint,prl,aps,showpacs]{revtex4}
\usepackage{graphicx}

\begin{document}

\title{Growth mechanism of nanocrystals in solution: ZnO, a case study.}
\author {Ranjani Viswanatha$^1$}
\author {Pralay K. Santra$^1$}
\author {Chandan Dasgupta$^{2~{\dagger}}$}
\author {D. D. Sarma$^{1,2,3~{\dagger}}$}
\email {sarma@sscu.iisc.ernet.in} \affiliation {$^1$Solid State and
Structural Chemistry Unit, Indian Institute of Science,
Bangalore-560012, India.} \affiliation{$^2$Centre for Condensed
Matter Theory, Department of Physics, Indian Institute of Science,
Bangalore-560012, India.} \affiliation {$^3$Centre for Advanced
Materials, Indian Association for the Cultivation of Science,
Kolkata 700032, India} \altaffiliation {Also at Jawaharlal Nehru
Centre for Advanced Scientific Research, Bangalore-560064, India}

\begin{abstract}
We investigate the mechanism of growth of nanocrystals from solution
using the case of ZnO. Spanning a wide range of values of the
parameters, such as the temperature and the reactant concentration,
that control the growth, our results establish a {\it qualitative}
departure from the widely accepted diffusion controlled coarsening
(Ostwald ripening) process quantified in terms of the
Lifshitz-Slyozov-Wagner theory.  Further, we show that these
experimental observations can be qualitatively and quantitatively
understood within a growth mechanism that is intermediate between
the two well-defined limits of diffusion control and kinetic
control.
\end{abstract}

\pacs{61.46.Df, 81.10.Dn, 81.10.Aj, 68.55.Ac}

\maketitle

The kinetics of growth and coarsening of clusters of a minority
phase in the background of a majority phase has been studied
extensively~\cite{onuki} for many years using analytic~\cite{bray},
computational~\cite{binder} and experimental~\cite{carlow,alkemper}
methods. In recent times, there has been renewed interest in
understanding the growth mechanisms in the nanometric
regime~\cite{alves,iwasawa,halas,narasimhan} to control
and manipulate various electronic properties of nano-scale systems.
However
these studies are concerned with the growth of a solid in a solid
medium and the growth of a solid in a solution, that constitutes an
important class of synthesis methods, is not extensively
investigated in the literature. This is probably due to the common,
but not fully substantiated, belief that in the synthesis from
solutions, the growth of the nanocrystals occurs via a diffusion
limited `Ostwald ripening' mechanism~\cite{searsonPRE,searson2}.
Though there exist some reports on non-Ostwald-ripening growth of
metal nanocrystals~\cite{cnr}, it is almost universally accepted
that the growth of semiconducting nanocrystals in the nanometric
regime proceeds via a diffusion controlled Ostwald ripening process,
as reported so far for TiO$_2$~\cite{tio2}, InAs and
CdSe~\cite{cdseinas} and ZnO~\cite{searson2} nanocrystal growths.
However, there exists a theoretical understanding of the growth
kinetics in both the well-defined limits of diffusion controlled
growth~\cite{bartels} and growth controlled by reaction kinetics
\cite{mitchell}. In this letter, we show that an analysis of our
experimental results for the growth of ZnO nanocrystals from
solution indicates that the growth process in this case is
qualitatively different from the expected Ostwald ripening behavior
and belongs to an intermediate regime between the two limiting
growth models, namely diffusion limited (Ostwald ripening) and
reaction limited growth. We have extended the well-known
Lifshitz-Slyozov-Wagner (LSW) theory~\cite{lifshitz,wagner} to
include the contribution from kinetically controlled growth. We find
that a treatment including the influences of both these processes on
the growth kinetics provides a consistent and quantitatively
accurate description of all experimental observations.

ZnO is a useful material for a wide range of applications, such as
solar cells, luminescent devices and chemical
sensors~\cite{appl,appl2}.
An intriguing aspect in the preparation of ZnO nanocrystals has been the
observation that the presence of a small amount of water in the
synthesis influences strongly the size of the
nanocrystals~\cite{meulankamp,dd_zno}.
In earlier
work~\cite{searson2} on the growth kinetics of ZnO
formation in water, it was concluded,
on the basis of the time dependence of the
average diameter, $d$, of ZnO nanocrystals, that
the growth follows the expected Ostwald ripening process. Besides
the well known $d \propto t^{1/3}$ law, Ostwald ripening,
characterized by a diffusion limited growth process, also requires
specific dependencies of the growth kinetics on the temperature and
the concentrations of the chemical reactants, as predicted by the
LSW theory~\cite{lifshitz,wagner}. Therefore, we have investigated
in detail the growth kinetics of ZnO nanocrystals in water to
establish the dependence of the average size on time, temperature
and reactant concentration. Our results show that though the average
diameter may be fitted to a cube-root of time dependence at long
time scales, the observed dependence on temperature and
reactant concentration is {\it qualitatively} different from what
would be expected from an Ostwald ripening process.

\begin{figure}[ht]
\vspace*{-0.5in}\includegraphics*[width=8.0cm]{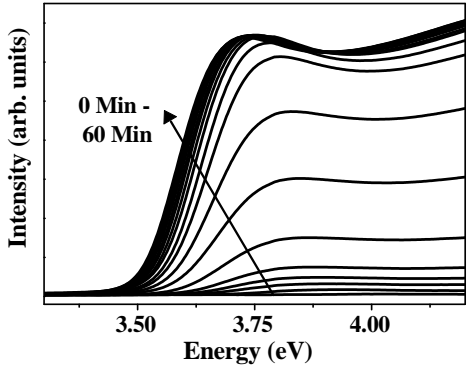}\vspace*{-2.25in}
\caption{\label{uvabs} (a) UV-absorption curves obtained at equal
intervals of time for a typical reaction carried out at 308 K with
100 mM of water.}
\end{figure}

A typical growth process involves adding 0.1~mmol of zinc acetate to
100~mL of 100-250~mM solution of double distilled water in
iso-propanol (i-PrOH) maintained in a water bath at the required
temperature (301-338K). In order to monitor the growth of ZnO
nanocrystals in real time during the growth process, we make use of
the {\it in-situ} time-resolved optical absorption spectra recorded
from the reaction mixtures. The well-known shifts of the bandgaps,
and consequently of the absorption edges with
size,~\cite{dd_zno,ddprb1,ddprb2,iiiv} provide a reliable way to
extract the average size and size distribution of the growing
nanocrystal assembly.~\cite{ddchemeur}  In order to confirm the
results obtained from the absorption technique, we have also carried
out transmission electron microscopy (TEM) at a few selected points
during the growth.
A typical set of optical-absorption spectra for a given
concentration of reactants and at a fixed temperature (308 K) is
shown in Fig.~\ref{uvabs}.  From the figure it is evident that the
bandgap shifts towards lower energy with increasing time, indicating
a systematic growth of nanocrystals.  Additionally, one can also
observe an increase in the absorption intensity with increasing
time, suggesting an increase in ZnO concentration with time.

\begin{figure}[ht]
\vspace*{-0.3in}\includegraphics*[width=8.0cm]{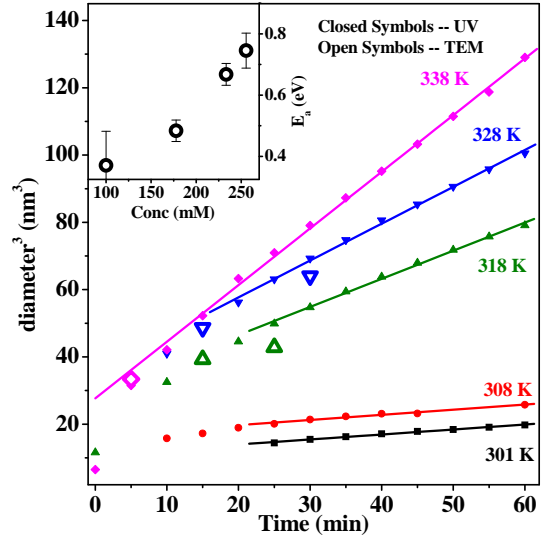}\vspace*{-1.6in}
\caption{\label{fig2} (color online) The cube of the average
diameter of ZnO nanocrystals shown as a function of time for
different temperatures at a fixed water concentration (100 mM).  The
inset shows the variation of the activation energy with the
concentration of water.}
\end{figure}

The rate law for diffusion limited growth or coarsening, often
termed~\cite{ostnew} as Ostwald ripening, was derived by Lifshitz
and Slyozov~\cite{lifshitz} and by Wagner.~\cite{wagner} According
to this theory the average diameter of the particles has a cube-root
dependence on time, following the relation $d^3 - d_0^3 = Kt$, where
$d$ is the average diameter at time $t$ and $d_0$ is the average
initial diameter of the nanocrystals. The rate constant $K$ is given
by $K = 8 \gamma D V_m^2 C_\infty / 9 R T$, where D is the diffusion
constant at temperature $T$, given by $D_0 \exp(-E_a/k_BT)$ ($E_a$
is the activation energy for diffusion), $V_m$ is the molar volume,
$\gamma$ is the surface energy and $C_\infty$ is the equilibrium
concentration at a flat surface. We show some typical variations of
$d^3~ vs.~ t$ for several temperatures in Fig.~\ref{fig2}, the
points obtained from TEM being shown as open symbols.  Though the
time-dependence of $d^3$ deviates from linearity at earlier times,
it indeed follows a linear relation at higher time-scales reasonably
well (as shown by the thick solid lines), suggesting a dominantly
diffusion limited growth in the long time limit. However, it should
be noted that an apparently linear dependence of $d^3$ on $t$,
especially only in the asymptotic limit, does not rigorously
establish the validity of the LSW theory, although this criterion
has been used~\cite{tio2,searson2} extensively in earlier studies of
growth of such particles, including the case of
ZnO. In general, $d^x$ as a function of $t$ may appear linear within
the experimental error limit for a wide range of $x$-values. We have
verified that the present results show acceptable linear behavior
for $x$-values ranging from 2.3 to 4.
Therefore, it
becomes necessary to explicitly verify the expected dependencies of
the rate constant $K$ on the temperature and the concentrations of
the reactants, which provide more sensitive and critical testing
grounds for the growth mechanism.  We have, therefore, analyzed
these dependencies in detail.

The dependence of the rate constant $K$ on the temperature, $T$,
arising primarily from the temperature-dependence of the diffusion
constant, $D$, should follow the activated form $K \propto
\exp(-E_a/k_BT)/T$.
We find that though the observed dependence of
$K$ on $T$ is reasonably well-described by this functional form at
higher concentrations of water, the fit is far from satisfactory for
lower water concentrations. Even more significantly, the activation
energy, $E_a$, obtained from the best-fit curves and plotted (open
circles) in the inset of Fig.~\ref{fig2} clearly shows a pronounced
dependence on the concentration of water.  While a
concentration-dependence of the activation energy may arise from
chemical diffusion at high water concentrations, the smallness of
the highest water concentration (0.46\%) employed here and the
difficulty in fitting the temperature-dependence of K even for a
fixed water concentration to the expected activated form suggest
that a purely diffusion-limited growth mechanism with a constant
activation energy assumed in the LSW theory cannot describe the
growth process of ZnO nanocrystals.

Growth of any nanocrystal via a solution route must be controlled
essentially by two processes.  One is the diffusion process of the
reactants to the surface of the growing crystallite, while the
second one is the reaction at the surface of the crystallite to
incorporate the reactant as a part of the growth process.  The
prevalent belief of the diffusion process being rate limiting leads
to the standard form of Ostwald ripening with a $d^3 \propto t$
dependence via the LSW theory.
The results presented here clearly establish that the details
of the growth kinetics, in particular its dependence on temperature
and the reactant concentration, invalidate the applicability of this
simplified approach, prompting us to probe the possible influence of
the surface reaction rate.  The reaction involves the dissociation
of zinc acetate, providing Zn$^{2+}$ ions. Hydroxyl ions are
produced in the solution from the dissociation of water. The
nanocrystals of ZnO comprise of tetrahedrally coordinated Zn and O
atoms and only the surface Zn atoms are terminated with a hydroxyl
ion instead of the oxygen ion. The growth of a nanocrystal occurs by
the dehydration of terminating OH$^{-}$ ions using the freely
available OH$^{-}$ ions in the solution. This is followed by the
capturing of Zn$^{2+}$ ions brought near the surface of the
nanocrystal by diffusion. The growth of the nanocrystal is further
continued by the Zn$^{2+}$ ion capturing an OH$^{-}$ ion and so on.
Thus the reaction, namely, $H_2O \rightleftharpoons H^+ + OH^-$,
$Zn^{2+} + 2OH^- \rightleftharpoons Zn(OH)_2 \rightleftharpoons ZnO
+ H_2O$, is controlled both by the diffusion of Zn$^{2+}$ ions and
the rate at which the reactions take place at the surface.  Hence
both these processes have to be taken into consideration in the
modeling of the growth process. This interpretation is qualitatively
supported by the experimental observations when we take into account
the increase of the dissociation constant of water by about two
orders of magnitude with increase in temperature, providing a large
number of OH$^-$ ions at higher temperatures. This increases the
rate of the reaction drastically at higher temperatures. Therefore,
it is expected that the growth process would shift towards a
diffusion controlled mechanism at higher temperatures. This is
entirely consistent with the results shown in Fig.~\ref{fig2},
showing that an improved conformity with $d^3 \propto t$ behavior
occurs systematically at earlier times at higher temperatures; a
similar trend is also seen at higher water concentrations for a
given temperature for the same reason.

In order to achieve a {\it quantitative} description, we note that
the rate of change of the radius $r$ of a growing cluster,
characterized by the diffusion constant $D$ and the reaction rate
constant $k_d$, is given by~\cite{sugimoto}
\begin{equation}
\frac{dr}{dt} = \frac{\kappa}{Tr^2} ~ \left(\frac {r/r_b-1}{1/D +
1/k_dr}\right), \label{rateeq}
\end{equation}
where, the constant $\kappa$ is given by $\kappa=2\gamma V_m^2
c_\infty/R$ and $r_b$ is the particle radius in equilibrium with the
solution. This equation implies that the reaction term is more
important for small $r$, which is consistent with our observation
(Fig.~\ref{fig2}) of more marked deviations from a purely diffusion
controlled growth at early times. Replacing $r$ by the average size
$d$, and assuming,~\cite{footnote} that the ratio of the average
radius and the equilibrium radius $r_b$ remains constant in time,
Eq.(\ref{rateeq}) can be integrated to obtain the relation $t = Bd^3
+ Cd^2 + const$,  with $B = K T \exp (E_a/k_BT)$ and $K \propto
1/(D_0\gamma V_m^2 c_\infty)$. The coefficient $C$ is of the form $C
\propto T/(k_d\gamma V_m^2 c_\infty)$. Thus, this equation not only
defines the dependence of the average diameter $d$ on time $t$, but
also separates out the diffusion and reaction terms. This equation
also yields the correct dependence of $d$ on $t$ in the two limiting
cases: $d^3 \propto t$ in diffusion limited growth ($D/k_dr \ll 1$),
and $d^2 \propto t$ in reaction limited growth ($D/k_dr \gg 1$).

\begin{figure}[ht]
\vspace*{-0.2in}\includegraphics*[width=8.0cm]{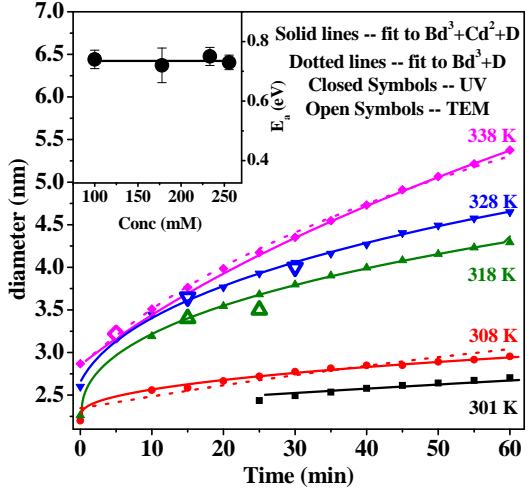}\vspace*{-1.82in}
\caption{\label{fig3} (color online) The average diameter of ZnO
nanocrystals shown as a function of time for different temperatures
at a fixed water concentration (100 mM). The solid (dashed) lines
show the best fits obtained over the entire data range using the
form $Bd^3 + C d^2 +D$ ($Bd^3 + D$). The inset shows the dependence
of the activation energy on water concentration.}
\end{figure}

We have used this expression to fit the experimentally observed
variation of the average diameter $d$ with time $t$, illustrated for
a given water concentration (100~mM) by thick solid lines through
experimental data obtained at different temperatures in
Fig.~\ref{fig3}. The remarkable goodness of fits over the entire
range of the data points, in contrast to fits obtained from the LSW
expression $t = Bd^3 + const$, illustrated with dotted lines in Fig.
3 for $T$ = 308 and 338~K only, provides a conclusive validation of
this description (Eq.~\ref{rateeq}). Further, we have obtained the
values of $B$ for different temperatures and different
concentrations of water from the fits. The expected temperature
dependence of the coefficient $B$ is given by $B \propto
T~\exp(E_a/kT)$.  The activation energy, $E_a$, obtained from the
least square fits to $B(T)$ is plotted as a function of water
concentration in the inset to Fig.~\ref{fig3}. In sharp contrast to
the results for the activation energy obtained earlier assuming only
a diffusion controlled growth (shown in the inset of
Fig.~\ref{fig2}), the new results show a concentration-independent
activation energy of 0.735 $\pm$ 0.007~eV, as expected.  This
provides a further validation of the proposed growth mechanism.

\begin{figure}[ht]
\vspace*{-0.95in}\includegraphics*[width=8.0cm]{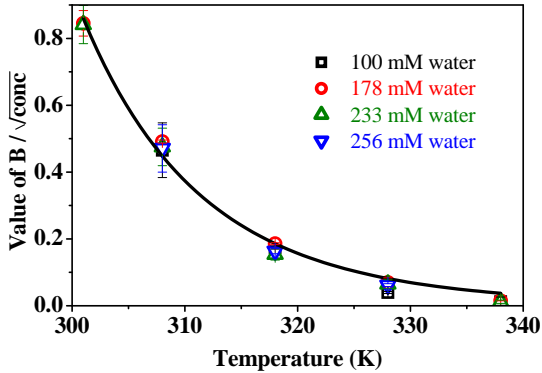}\vspace*{-1.7in}
\caption{\label{fig4} (color online) The constant $B$ scaled by the
square root of water concentration, as a function of temperature for
different water concentrations.}
\end{figure}

We find that the values of $B$ obtained at a fixed temperature for
various water concentrations are proportional to the square root of
the water concentration, evidenced by the collapse of $B/(water~
concentration)^{1/2}$ vs $T$ plots into a universal curve in
Fig.~\ref{fig4}.  While we still do not have a rigorous explanation
for this interesting behavior, such a collapse of the $B$ values may
be understood qualitatively in the following way. It is known that
the diffusion constant $D_0$, molar volume $V_m$, surface energy
$\gamma$ and the equilibrium concentration $c_\infty$ at a flat
interface cannot depend on the concentration of water, at least for
the small changes in water concentration used here. However,
the growth flux depends on the rate and the extent of reactions
occurring near the surface which, in turn, depend on the
concentration of the OH$^-$ ions; assuming the concentration of
H$^+$ and OH$^-$ ions to be the same, the concentration of the
OH$^-$ should be proportional to the square root of the water
concentration.  Thus, the observed dependence of $B$ on water
concentration is consistent with the local chemistry occurring at
the surface.

In conclusion, we show that a linear dependence of the cube of the
average diameter on time is not a critical test to determine the
growth mechanism.
If the
present data set is analyzed in terms of the $d^3 \propto t$
relationship predicted by the LSW theory in the diffusion controlled
regime, the estimated activation energy for diffusion shows an
unphysical dependence on the concentration of water.  This implies a
clear departure from
the diffusion-limited Ostwald ripening
process.  However the expression obtained for the time dependence of
the average diameter for growth controlled by both the rate of
diffusion and the rate of reaction at the surface provides good
agreement with experimental results over the entire range of time,
temperature and concentration. In addition, this approach provides
an estimate for the activation energy for the diffusion process that
is independent of water concentration.
These observations firmly
establish that the mechanism of growth lies in the intermediate
regime of diffusion and kinetically controlled growth processes.

The work is supported by the Department of Science and Technology,
Government of India. We thank Dr. Zhenyu Zhang for his constructive
criticisms which improved the quality of the paper.



\begin{thebibliography}{99}

\bibitem{onuki} A. Onuki, {\it Phase Transition Dynamics}
(Cambridge University Press, Cambridge, 2002).

\bibitem{bray} A. J. Bray, Adv. Phys. {\bf 43}, 357 (1994).

\bibitem{binder} K. Binder and P. Fratzl, in {\it Phase Transformations
in Materials}, edited by G. Kostorz (Wiley-VCH, Weinheim, New York, 2001).

\bibitem{carlow} G. R. Carlow and M. Zinke-Allmang, Phys. Rev Lett. {\bf 78},
4601 (1997).

\bibitem{alkemper} J. Alkemper, V. A. Snyder, N. Akaiwa, and P. W. Voorhees
Phys. Rev. Lett. {\bf 82}, 2725 (1999).


\bibitem{narasimhan}
P. Gangopadhyay {\it et al.}, Phys. Rev. Lett., {\bf 94}, 047403
(2005).

\bibitem{iwasawa}
S. Takakusagi, K. Fukui, R. Tero, F. Nariyuki and Y. Iwasawa, Phys.
Rev. Lett., {\bf 91}, 066102 (2003).

\bibitem{alves}
A. F. Craievich, G. Kellermann, L. C. Barbosa and O. L. Alves, Phys.
Rev. Lett., {\bf 89}, 235503 (2002).

\bibitem{halas}
R. D. Averitt, D. Sarkar and N. J. Halas, Phys. Rev. Lett., {\bf
78}, 4217 (1997).

\bibitem{searsonPRE} G. Oskam, Z. Hu, R.L. Penn, N. Pesika, P.C.
Searson, Phys. Rev. E, {\bf 66}, 11403 (2002).

\bibitem{searson2}
Z. Hu, D. J. Escamilla Ramirez, B. E. Heredia Cervera, G. Oskam and
P. C. Searson, J. Phys. Chem. B, {\bf 109}, 11209 (2005).

\bibitem{cnr}
R. Seshadri, G. N. Subbanna, V. Vijayakrishnan, G. U. Kulkarni, G.
Ananthakrishna and C. N. R. Rao, J. Phys. Chem., {\bf 99}, 5639
(1995).

\bibitem{tio2}
G. Oskam, A. Nellore, R. L. Penn and P. C. Searson, J. Phys. Chem.
B, {\bf 107}, 1734 (2003).

\bibitem{cdseinas}
X. Peng, J. Wickham and A. P. Alivisatos, J. Am. Chem. Soc., {\bf
120}, 5343 (1998).

\bibitem{bartels}
J. Bartels, U. Lembke, R. Pascova, J. Schmelzer and I. Gutzow, J.
Non-Cryst. Sol., {\bf 136}, 181 (1991).

\bibitem{mitchell}
D. Du, D. J. Srolovitz, M. E. Coltrin and C. C. Mitchell, Phys. Rev.
Lett., {\bf 95}, 155503 (2005).

\bibitem{lifshitz}
I. M. Lifshitz and V. V. Slyozov, J. Phys. Chem. Solids, {\bf 19},
35 (1961).

\bibitem{wagner}
C. Wagner, Z. Elektrochem., {\bf 65}, 581 (1961).

\bibitem{appl}
\emph{Thin film solar cells} edited by K. L. Chopra, S. R. Das,
(Plenum, New York, 1983)

\bibitem{appl2}
S. Hingorani, V. Pillai, P. Kumar, M. S. Multani and D. O. Shah,
Mater. Res. Bull., {\bf 28}, 1303 (1993).

\bibitem{meulankamp}
E. A. Meulenkamp, J. Phys. Chem. B, {\bf 102}, 5566 (1998).

\bibitem{dd_zno}
R. Viswanatha, S. Sapra, B. Satpati, P. V. Satyam, B. N. Dev and D.
D. Sarma, J. Mater. Chem, {\bf 14}, 661 (2004).

\bibitem{ddprb1}
S. Sapra, N. Shanthi and D. D. Sarma, Phys. Rev. B, {\bf 66}, 205202
(2002).

\bibitem{ddprb2}
S. Sapra and D. D. Sarma, Phys. Rev. B, {\bf 69}, 125304 (2004).

\bibitem{iiiv}
R. Viswanatha, S. Sapra, T. Saha-Dasgupta and D. D. Sarma, Phys.
Rev. B, {\bf 72}, 045333 (2005).

\bibitem{ddchemeur}
R. Viswanatha and D. D. Sarma, Chem. Eur. J., {\bf 12}, 180 (2006).

\bibitem{ostnew}
R. Finsy, Langmuir, {\bf 20}, 2975 (2004).

\bibitem{sugimoto}
T. Sugimoto, Adv. Coll. Inter. Sci., {\bf 28}, 65 (1987).

\bibitem{footnote}
This is true in both the limiting cases of purely diffusion
controlled~\cite{lifshitz} and purely reaction
controlled~\cite{sugimoto} growth.

\end{thebibliography}
\end{document}